\documentclass[reprint,showpacs,showkeys,english,aps,prb]{revtex4-1}
\usepackage[T1]{fontenc}
\usepackage{subfigure}
\usepackage[latin1]{inputenc}
\usepackage{graphicx}
\usepackage{epsfig}
\usepackage{prettyref}
\usepackage{babel}

\begin{document}
\title{Heterogeneous nucleation and metal-insulator transition in epitaxial films of NdNiO$_3$}

\author{Devendra Kumar}

\email{deven@iitk.ac.in}

\affiliation{Department of Physics, Indian Institute of Technology
Kanpur 208016, India}

\author{K.P. Rajeev}

\email{kpraj@iitk.ac.in}

\affiliation{Department of Physics, Indian Institute of Technology
Kanpur 208016, India}

\author{A. K. Kushwaha}

\affiliation{Department of Physics, Indian Institute of Technology
Kanpur 208016, India}

\author{R. C. Budhani}

\affiliation{Department of Physics, Indian Institute of Technology
Kanpur 208016, India}

\begin{abstract}
We have investigated the temperature driven first order
metal-insulator transition in thin films of NdNiO$_3$ and have
compared it with the bulk behavior. The M-I transition of thin
films is sensitive to epitaxial strain and a partial relaxation of
epitaxial strain creates an inhomogeneous strain field  in the
films which broadens the M-I transition. Both the thin film and
the bulk samples exhibit non equilibrium features in the
transition regime which are attributed to the presence of high
temperature metallic phases in their supercooled state. The degree
of supercooling in the thin films is found to be much smaller than
in the bulk which suggests that the metal insulator transition in
the thin film occurs through heterogeneous nucleation.
\end{abstract}



\maketitle
\section{Introduction}
The study of first order metal-insulator (M-I) transition in
transition metal oxides has been an active area of research for
several decades now because of fundamental interest and exciting
physical phenomena associated with these systems, for example
hysteresis, slow dynamics, phase separation, colossal
magnetoresistance etc.\cite{Imada, Dagotto, Morin, Granados}
Whereas the presence of dynamical features has been reported in
many early studies,\cite{Morin, Granados} an attempt to understand
them and their dependence on the parameters controlling the M-I
transition and their relationship with other associated phenomena
such as hysteresis, phase separation  has only started around the
beginning of this century.\cite{Chaddah, Chaddah1, Chaddah2,
Devendra, Devendra1} In these oxides, the family of rare earth
nickelates (RNiO$_3$, R$\neq$La) exhibits an M-I transition which
can be tuned by application of internal or external pressure or
strain.\cite{Medarde, Obradors, Canfield, Kozuka, Staub, Ashutosh,
Novojilov, Gatalan1, Gatalan, Kaura, Eguchi, Conchon, Ambrosinia,
Capon} The M-I transition, the hysteresis and their connection to
slow dynamics have been studied in detail in bulk NdNiO$_3$ and
PrNiO$_3$.\cite{Granados, Devendra, Devendra1} In
Sm$_x$Nd$_{1-x}$NiO$_3$ and Nd$_{0.7}$Eu$_{0.3}$NiO$_3$ thin films
it has been observed that an increase in the film substrate
lattice mismatch strain broadens the M-I transition and reduces
the associated hysteresis.\cite{Ambrosinia, Capon} The physics of
such behavior has not been explained so far. In this work we make
an attempt to understand the effect of substrate strain on M-I
transition and the differences in the M-I transition of the bulk
and the thin films of NdNiO$_3$ by performing a comparative study
of a few relevant features of these systems.

The existence of dynamical features in a first order phase
transition (FOPT) system can be understood on the basis of the
Landau formulation of mean field theory. In this theory, the free
energy of the system is expressed in terms of an order parameter
in the vicinity of the thermodynamic transition. In an FOPT system
that undergoes a metal to insulator transition on cooling, at
$T=T_{MI}$, the free energy of the system has two minima of equal
depth separated by an energy barrier. One of the minima occurs at
zero value of the order parameter and this corresponds to the high
temperature metallic state and the other minimum occurs at a
finite value of the order parameter and it corresponds to the low
temperature insulating state.\cite{Chaikin}  See figure
\ref{fig:free energy}. On lowering the temperature further, the
energy of the insulating state decreases and it becomes the stable
state of the system while the metallic state exists as a
supercooled (SC) metastable state. The high temperature metallic
phase present in its SC state will remain in
 that state till it gains enough energy to overcome the free energy  barrier
($U$) that separates it from the stable insulating
state.\cite{Chaikin, Chaddah} For $T<T_{MI}$ the energy barrier
($U$) is the energy required for the formation of a critical
nucleus of the stable phase inside the SC phase. Once a critical
nucleus is formed or in other words when the energy barrier $U$ is
crossed, the rest of the SC phase will progressively transform or
switch to the stable insulating state. At constant pressure and
temperature, the energy required for the nucleation of a critical
nucleus of volume $V$ and surface area $S$ is given as
\cite{Porter}:
\begin{equation}
U=-V\Delta G_{V} + S\gamma_{S} + V\Delta
G_{Strain}\label{eq:nuc-eqn}
\end{equation}
where  $\Delta G_{V}$ is the difference in the Gibb's free energy
per unit volume of SC metallic and stable insulating states,
$\gamma_{S}$ is the energy required per unit area for the
formation of a metal-insulator interface and  $\Delta  G_{Strain}$
is the misfit strain energy per unit volume. At $T=T_{MI}$,
$\Delta G_{V}$ is zero. So the  critical nucleus at this
temperature is essentially the complete metallic phase and  $V$
and $S$ in equation \ref{eq:nuc-eqn} are equal to the volume and
surface area of the metallic phase.\cite{endnote2} Below $T_{MI}$,
$\Delta G_{V}$ increases monotonically on lowering the temperature
which decreases $U$ and the size of critical nucleus. Thus the
energy of nucleation $U$ is maximum at $T=T_{MI}$ and decreases
monotonically on lowering the temperature below $T_{MI}$. At a
critical temperature $T^*$, which is known as the limiting
temperature of metastability, $U$ vanishes.\cite{Chaikin} Now the
SC metallic phase is free to switch to the stable insulating state
and if the dynamics of the system is not frozen (i.e $T > T_g$,
$T_g$ being the temperature of kinetic arrest), then these SC
regions will switch to the insulating state.\cite{Chaddah1,
Devendra} The presence of preferential sites for heterogeneous
nucleation such as defects, interphase boundaries and free
surfaces decreases $\gamma_S$ and $\Delta G_{Strain}$ which in
turn could decrease $U$ and the degree of maximum possible
supercooling ($T_{MI}-T^*$).\cite{Porter} In systems where such
preferential sites for nucleation are missing, the SC phase will
transform through homogeneous nucleation where usually $U$ is
large and consequently the extent of maximum supercooling
$(T_{MI}-T^*)$ is also high.

  In crystals quenched disorder such as lattice distortion, defects, strain fields
and non-stoichiometry are usually induced during crystal growth.
The presence of quenched disorder  causes a local variation of the
$T_{MI}$ resulting in a broadening of the otherwise sharp first
order M-I transition.\cite{Imry, Soibel, Chaddah2, Dagotto1} The
local variation of $T_{MI}$ will also cause a local variation of
$T^*$. Thus in a real system, while cooling, below the M-I phase
transition temperature, we will have a collection of regions with
definite $T_{MI}$ and $T^*$ present either in metallic or
insulating state. These regions switch from one state to the other
as a single entity and we will refer to them as switchable regions
(SR). In this set of SR's there would be a fraction which will
have $T^* < T <T_{MI}$. Of these, the SR's which are in their
metallic state, with $T_g < T$,  switch over to the insulating
state stochastically giving rise to time dependence in physical
properties of non-glassy and non-spin-glass type FOPT systems.
This in turn results in a hysteresis between cooling and heating
data.\cite{Devendra, Devendra1}

In this paper we report time and temperature dependent resistivity
measurements on thin films of the FOPT system NdNiO$_3$ and has
compared the results with the bulk. Our results show that the M-I
transition in the bulk is associated with a large hysteresis and
strong dynamical effects while  in thin films it has a small
hysteresis and weak dynamical effects. We have discussed these
results using the concepts of supercooling and nucleation which
suggest that the energy ($U$) required for the nucleation of a
critical insulating nucleus inside the metallic phase is much
smaller for thin films compared to the bulk. This indicates that
the M-I transition in the bulk occurs predominantly through
homogeneous nucleation while the M-I transition in the films
occurs predominantly through heterogeneous nucleation.

\begin{figure}[!t]
\begin{centering}
\includegraphics[width=0.6\columnwidth]{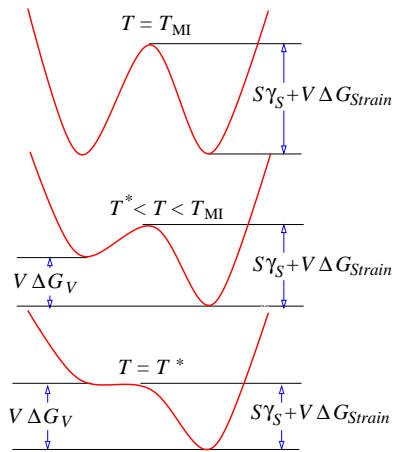}
\par\end{centering}
\caption{(Color Online) Dependence of free energy on temperature
in the vicinity of a first order metal-insulator transition. At T
= $T_{MI}$ the metallic and insulating states have equal free
energy and they are separated by an energy barrier U(=$S\gamma_{S}
+ V\Delta G_{Strain}$). On lowering the temperature below $T_{MI}$
the free energy of insulating state decreases. This results in a
decrease in energy barrier U which eventually vanishes at a
temperature T = $T^*$ (See text for details).} \label{fig:free
energy}
\end{figure}

\section{Experimental Details and Results}
\subsection{Thin film deposition and structural analysis}\label{sec:deposition}
\begin{figure}[!t]
\begin{centering}
\includegraphics[width=1\columnwidth]{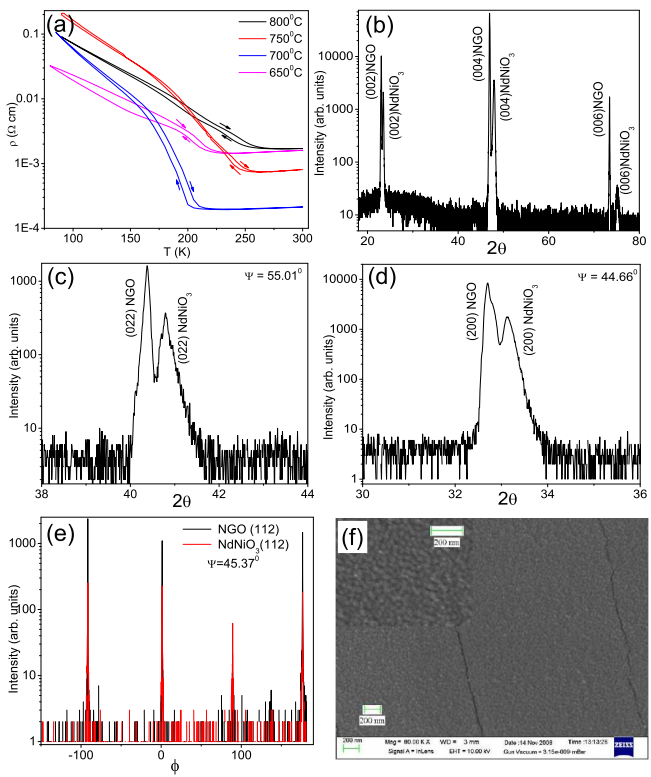}
\end{centering}
\caption{(Color Online) (a) Temperature dependence of resistivity
of NdNiO$_3$ thin films deposited at 650-800$^0$C. (b)  X-ray
$\theta-2\theta$ diffraction pattern  taken with Cu K$\alpha$
radiation for the 700$^0$C film , (c) asymmetric $\theta-2\theta$
plot of (220) plane, (d) asymmetric $\theta-2\theta$ plot of (002)
plane, (e) and  $\phi$ san plot of (112)plane of NdNiO$_3$ thin
film (red line) and NGO substrate (black line). $\Psi$ is the tilt
angle of the surface normal of the film. (f) Scanning electron
micrograph of NdNiO$_3$ thin film grown at 700$^0$C. The inset
shows a higher resolution image of the same surface.}
\label{fig:xray}
\end{figure}

High quality epitaxial films of NdNiO$_3$ of thickness~200 nm were
grown by pulsed laser deposition on \{001\} oriented single
crystal NdGaO$_3$ (NGO) substrates.  An NdNiO$_3$ pellet prepared
by a low temperature method,\cite{Vassiliou} and sintered at 1100
$^0C$ for 12 hours in an oxygen atmosphere, was used as the target
for the thin film deposition. The energy density and pulse
frequency for the deposition were 1.9 J/cm$^2$ and 10 Hz
respectively. The films were deposited at 0.3 mbar oxygen pressure
and the substrate temperature was varied between 650-800$^0$C.
After the deposition the sample is cooled slowly at 1 bar O$_2$
pressure. These films were characterized by resistivity
measurements and the results are shown in figure
\ref{fig:xray}(a). We found that the films are sensitive to the
deposition temperature and that the films deposited at 700~$^0$C
have the best features. It has the smallest room temperature
resistivity and shows the maximum relative change in resistivity
on going from metallic to insulating state. We have used this film
for the rest of the study. The sensitivity of NdNiO$_3$ thin films
to the deposition temperature has been reported earlier
also.\cite{Kaura, Eguchi} Our thin film samples were also
characterized by X-ray diffraction using CuK$\alpha$ radiation on
a PANalytical's X'Pert PRO diffractometer and by scanning electron
microscopy using a SUPRA 40VP FESEM and the data for the film
deposited at 700~$^0$C is shown in figure \ref{fig:xray}(b)-(f).
The $\theta- 2\theta$ X-ray diffraction of NdNiO$_3$ film  has
peaks only corresponding to \{001\} plane of NGO substrate which
indicates that the NdNiO$_3$ film is either highly <001> oriented
or grown epitaxially on NGO substrate with out of pane lattice
parameter $c_{NGO}$=7.710 {\AA} and $c_{film}$=7.573 {\AA}. The
FWHM of the (004) peak of the thin films is 0.31 degree which is
higher than that of bulk samples (0.18 degree). This gives the
average grain size of thin films as 60~nm which is smaller than
that of bulk sample (132~nm).

The in plane lattice  parameter of thin films were obtained by the
asymmetric $\theta-2\theta$ scan of the (200) and (022) planes.
Table \ref{tab:lattice-parameters} shows the lattice parameters
$a$, $b$, $c$, Ni-O-Ni bond angle $\Theta$ and unit cell volume
$V$ for thin films and the bulk sample. A comparison of the
lattice parameters $a$, $b$, and $c$ show that the NdNiO$_3$ thin
films are under in-plane tensile strain and out-of-plane
compressive stress because of NdGaO$_3$ substrate. The increase in
the in-plane lattice parameters $a$ and $b$ of thin films should
have decreased the out-of-plane lattice parameter $c$ in such a
way that the unit cell volume remains conserved. But this is not
the case and the thin films have a higher unit cell volume
compared to bulk NdNiO$_3$. The unit cell volume of the
NdNiO$_{3}$ depends on the oxygen vacancies and it increases on
increasing the vacancies.\cite{Tiwari, Nikulina} This suggest that
in thin films a partial stress relaxation has happened by the
formation of oxygen vacancies. Similar chemical stress relaxation
has been observed in SmNiO$_3$ films deposited on SrTiO$_3$
substrates.\cite{Conchon} The X-ray $\phi$ scan measurement were
performed to check the epitaxy of the thin films. In figure
\ref{fig:xray}(e) we show the $\phi$ scan plot of (112) plane  of
NdNiO$_3$ film deposited on (001) oriented NGO substrate. We get
four equal spaced peaks for NdNiO$_3$ which are aligned with that
of  NGO. This clearly indicates the epitaxial growth of NdNiO$_3$
on NGO substrate. Figure \ref{fig:xray}(f) shows the scanning
electron micrograph of the film surface.  The film surface is
formed of tiny (001) oriented grains. The size of these grains is
much smaller than that of bulk samples.\cite{endnote} The surface
of the film shows few nano-scale line cracks. These cracks usually
form due to relaxation of the film-substrate lattice mismatch
strain and the differential contraction of the film and substrate
during the cooling process after the thin film
deposition.\cite{Freund, Kazuki, Copetti, Tian}
\begin{table}[]
\begin{centering}
\begin{tabular}{|c|c|c|c|c|c|c|}
\hline \
 & a ({\AA}) & b ({\AA}) & c ({\AA}) & $\Theta$ (deg.) & $V$ ({\AA}$^3$)\tabularnewline
\hline \hline Bulk & 5.3888 & 5.3845 & 7.6127 & 157.5 & 220.89
\tabularnewline \hline  Thin film & 5.406 & 5.447 & 7.573 & 155.4
& 223.00\tabularnewline \hline NdGaO$_3$ & 5.431 & 5.499 & 7.710 &
- & 230.26\tabularnewline \hline
\end{tabular}
\par\end{centering}
\caption{\label{tab:lattice-parameters}Lattice parameters a, b, c,
Ni-O-Ni bond angle $\Theta$ and  unit cell volume $V$ of bulk
NdNiO$_3$(from Ref.\ \onlinecite{P.Lcorre}), thin films of
NdNiO$_3$(calculated from the X-ray data) and the NdGaO$_3$
substrate. The Ni-O-Ni bond angle of thin film was calculated from
Hayashi's formula.\cite{Hayashi}}
\end{table}

\subsection{Transport measurements and results}
\begin{figure}[!t]
\begin{centering}
\includegraphics[width=0.9\columnwidth]{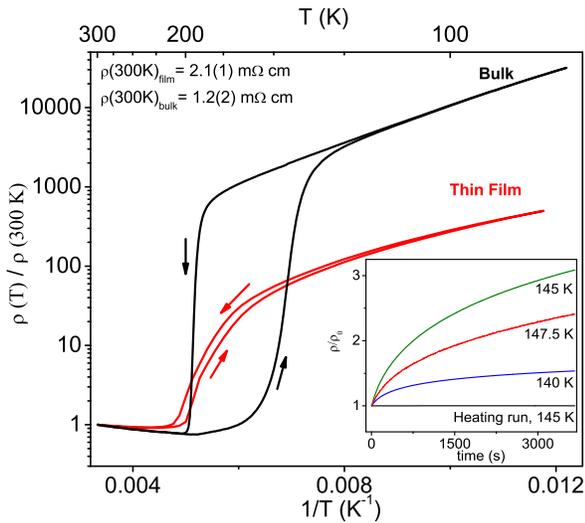}
\par\end{centering}
\caption{(Color online) $\rho (T)/\rho (300 K)$ versus $1/T$ plot
for thin film and bulk NdNiO$_3$. Inset shows the time dependence
of resistivity of bulk NdNiO$_3$ at 140~K, 145~K and 147.5~K while
cooling, and at 145~K while heating, for a period of 1h. The
resistivity data of bulk NdNiO$_3$ has been reported
earlier.\cite{Devendra}} \label{fig:R vs T}
\end{figure}

Figure \ref{fig:R vs T} shows the normalized electrical
resistivity of our best NdNiO$_3$ thin film  as a function of
temperature. The rate of temperature variation was fixed at
2~Kmin$^{-1}$ for all the experiments. The resistivity is multiple
valued, the cooling and heating data differ from each other and
form a hysteresis loop. The resistivity plot indicates that the
NdNiO$_3$ thin film begins to undergo a first order
metal-insulator transition at 215~K. In the same figure we have
also plotted the resistivity of bulk NdNiO$_3$ for easy
comparison. We can see that the beginning of the M-I transition in
thin films is about 15~K higher than that of bulk sample. The M-I
transition temperature, metallic resistivity and the ratio of
metallic to insulating resistivity of NdNiO$_3$ thin films depends
on the film thickness, choice of substrate and the deposition
condition.\cite{Ashutosh, Gatalan1, Gatalan, Kozuka, Conchon,
Kaura, Eguchi} For NdNiO$_3$ thin films deposited on (001)
oriented NGO substrate, the higher value of M-I transition
temperature than the bulk has also been reported earlier and it is
attributed to in plane tensile strain and out of plane compressive
stress present in these films.\cite{Eguchi} The epitaxial strain
of the films decreases the Ni-O-Ni superexchange angle of the
films compared to the bulk sample. See table
\ref{tab:lattice-parameters}. This results in a  bandwidth
narrowing which subsequently increases the M-I transition
temperature of the films. Similar effects has been observed in the
bulk samples where an increase in distortion (by decreasing the
rare earth ionic radius) increases the $T_{MI}$.\cite{Medarde} The
higher value of M-I transition temperature in thin films and the
comparison of lattice parameters show that the NdNiO$_3$ thin
films are more distorted compared to the bulk.

 At room temperature the resistivity of the thin film is 2.1(1) m$\Omega$cm which is higher
than the resistivity of bulk sample 1.2(2) m$\Omega$cm. The ratio
of insulating to metallic resistivity $\rho (80 K)/\rho (300K)$ of
thin film is about 60 times lower than that of bulk NdNiO$_3$. See
figure \ref{fig:R vs T}. The higher metallic resistivity of thin
films may be attributed to their smaller grain size and also to
the defects (line cracks and vacancies) formed by partial stress
relaxation.\cite{Fu, Gupta, Kazuki, Copetti, Tian, Conchon,
Nikulina} The large density of grain boundaries and defects
enhances the electron scattering which increases the resistivity
of the metallic state. On the other hand, in the insulating state
the higher density of grain boundaries and vacancies create defect
levels in the band gap which lowers the insulating resistivity of
the films.\cite{Nikulina, Shockley, Victor, Mckeena, Brassard}
Similar effect of grain size dependence has also been observed in
VO$_2$ thin films.\cite{Brassard} In the cooling run the M-I
transition of bulk NdNiO$_3$ is centered at 155~K with a spread of
about 90~K while in the heating run the M-I transition is
relatively sharp and has a width of about 10~K. Below 110~K the
heating and cooling data merge and the log$\rho$ versus $1/T$ plot
is linear. In the case of thin film we note that the M-I
transition is broader and it spreads over a temperature range of
about 120~K. In contrast to bulk, the width of M-I transition for
thin films is nearly the same for cooling and heating runs. We
further note that the amount of hysteresis in the films is much
less than that of the bulk system. Below 95 K the heating and
cooling data merge and the log$\rho$ versus $1/T$ plot is linear.

\begin{figure}[!t]
\begin{centering}
\includegraphics[width=0.9\columnwidth]{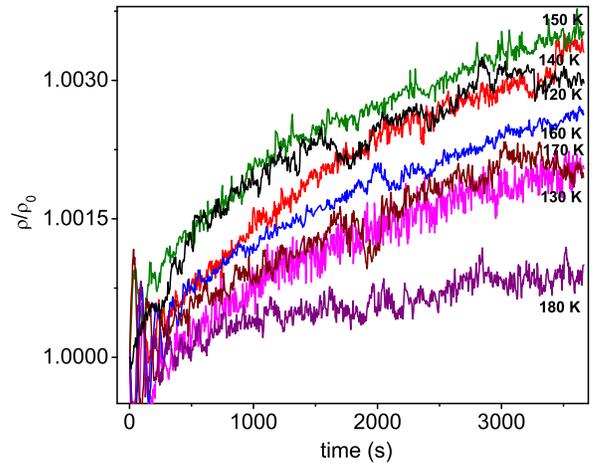}
\par\end{centering}
\caption{(Color online) Time dependence of resistivity for
NdNiO$_3$ thin film while cooling, at various temperatures in the
range of 180-120~K, for a period of 1h.} \label{fig:time dep}
\end{figure}

Figure \ref{fig:time dep} exhibits a subset of the time dependent
resistivity data for the thin films in the cooling run. The time
dependent resistivity data was recorded using the following
protocol. The sample was cooled from 250~K (>$T_{MI}$) to the
temperature of interest and the resistivity of the sample was
recorded as a function of time. \cite{Endnote} The data is
presented in the form of $\rho(t)/\rho(t=0)$ so that the values
are normalized to unity at t=0 for easy comparison. We found that
below 200~K the resistivity of the film increases with time. The
presence of time dependence in resistivity indicates that the
system is not in thermodynamic equilibrium. These curves were
fitted to the stretched exponential function
\begin{equation}
\rho(t)=\rho_{0}+\rho_{1}\left(1-e^{-\left(\frac{t}{\tau}\right)^{\gamma}}\right)\label{eq:Stretched-Exponential}
\end{equation}
where $\rho_{0}$, $\rho_{1}$, $\tau$ and $\gamma$ are fit
parameters. The thin film time dependent data is noisy compared to
the bulk samples and the quality of fitting is not as good as in
bulk samples.  See Table \ref{tab:Fit-parameters}. A maximum
relative increase in resistivity of about 0.35\,\% occurs at
around 150~K.\cite{Endnote1} The relative increase in the
resistivity ($\rho/\rho_0$)  of the thin films of NdNiO$_3$
(maximum 0.35\,\% in one hour)  is much less than that observed in
polycrystalline NdNiO$_3$ (maximum 200\,\% in one hour at 145~K .
See inset of figure \ref{fig:R vs T} and Ref.\
\onlinecite{Devendra}). No detectable time dependence was observed
below 95~K and in the subsequent heating run. The fitting
parameter $\gamma$ of thin films ($\gamma \approx0.7$) is more
close to 1 compared to bulk samples ($\gamma \approx0.5$), which
indicates that at the temperature of relaxation the SR's of thin
films have a narrow distribution of $T^*$ in comparison to bulk.

\begin{table}
\begin{centering}
\begin{tabular}{|c|c|c|c|c|c|c|}
\hline \# & T(K) & $\rho_{1}/\rho_{0}$ & $\tau$ ($10^{3}$s) &
$\gamma$ & $\chi^{2}/DOF$ & $R^{2}$\tabularnewline \hline \hline 1
& 120.0 & 0.0043(2) & 1.7(1) & 0.64(3) & 4.44 &
0.97309\tabularnewline \hline 2 & 130.0 & 0.0038(4) & 2.7(6) &
0.75(6) & 6.63 & 0.94336\tabularnewline \hline 3 & 140.0 &
0.0055(3) & 3.3(3) & 0.78(3) & 1.86 & 0.98603\tabularnewline
\hline 4 & 150.0 & 0.0043(3) & 2.5(2) & 0.79(3) & 1.66 &
0.98584\tabularnewline \hline 5 & 160.0 & 0.006(3) & 6(7) &
0.49(8) & 2.54 & 0.93284\tabularnewline \hline
\end{tabular}
\par\end{centering}

\caption{\label{tab:Fit-parameters}Fit parameters for the time
dependence data shown in Figure \ref{fig:time dep}. The degrees of
freedom of the fits $DOF\approx1000$. For 160~K, the error in
$\tau$ is higher than $\tau$ itself. Above 160~K, the data is too
noisy and the fitting does not yield proper results.}
\end{table}
\section{Discussion}
\subsection{Behavior of electrical resistivity and nucleation of insulating phases}
As shown in Figure \ref{fig:R vs T} the M-I transition for bulk
NdNiO$_3$ in cooling run starts from 200~K, and on lowering the
temperature more and more regions (SR's) of the sample transform
to the insulating state. Below 110~K, cooling and heating data
merge and it follows the band gap model of insulators which
indicates that most of the SR's have completed their transition to
the insulating state. The resistivity of the system exhibits a
large time dependence (maximum 200\,\%) in the cooling run which
suggest that a large fraction of SR's are present below their
respective $T_{MI}$ in their SC metastable state. The time
dependent effects persists down to 110~K which shows that $T^*$ of
SR's  are distributed in the range of 200~K to
110~K.\cite{Devendra, Devendra1} In the heating run the M-I
transition is narrow, is centered  around 195~K and has a width of
about 10~K. The time dependence of resistivity in the heating run
is negligible compared to the cooling run. This rules out the
possibility of superheating and suggests that the temperature
where an SR changes from insulating to metallic state in a heating
run is the $T_{MI}$ of that SR. The broadening of M-I transition
in the heating run is due to disorder induced variation of local
$T_{MI}$. So in the bulk sample, local $T_{MI}$ has a variation of
200-190~K while the $T^*$ has a variation from 200-110~K. The
existence of metallic SR's in their SC state much below their
respective $T_{MI}$ (a high degree of supercooling) suggests that
in cooling runs most of the SR's transform from metallic to
insulating state through homogeneous nucleation.

In the case of thin films the M-I transition is much broader than
in the bulk, it starts from 215~K and occurs over a temperature
range of about 120~K. The higher broadening of the M-I transition
in the films is possibly related to epitaxial strain of the films.
During the epitaxial growth of the films if the elastic energy of
the film due to substrate lattice mismatch strain exceeds the
critical energy required for the creation of defects, then the
film partially relaxes its elastic energy by formation of
defects.\cite{Freund} The defects can be of mechanical(eg. cracks,
stacking faults) or chemical (eg. creation of oxygen vacancies) in
nature.\cite{Kazuki, Copetti, Tian, Conchon} The elastic energy
depends on the amount of lattice mismatch and the thickness of the
film and for a film-substrate combination, above a critical
thickness or strain, defects will be formed to reduce the elastic
energy. The variation of epitaxial strain across the thickness of
the film is not uniform, and in general it is pronounced close to
the thin film substrate interface and relaxes on moving towards
the free surface.\cite{Conchon, Boulle} Thus the formation of
defects causes a variation of strain field inside the films. This
will results in the local variation of $T_{MI}$ inside the films
and it will broaden the metal to insulator transition of the
films. This effect has been seen experimentally in (a) thickness
dependent studies of NdNiO$_3$, SmNiO$_3$ and VO$_2$ thin films
where the thinner films exhibits a sharper M-I transition compared
to thick films.\cite{Gatalan, Conchon, Kazuki} (b)
Sm$_x$Nd$_{1-x}$NiO$_3$ thin films where the increase in Sm
content results in broader M-I transition.\cite{Ambrosinia} In our
thin films the occurrence of  partial stress relaxation by
formation of defects has already been discussed in section
\ref{sec:deposition}, and it shows itself in the form of line
cracks and increase in unit cell volume. See figure
\ref{fig:xray}(f) and table \ref{tab:lattice-parameters}. This is
a pointer to the presence of inhomogeneous strain fields in the
film which causes a local variation of $T_{MI}$ and a broad M-I
transition.

  The broadness of the M-I transition and the nature of $\rho$ vs
$1/T$ curve of the film are nearly the same for the heating and
cooling runs. This means that for most of the SR's the
temperatures where it transforms from metallic to insulating state
in cooling run and from insulating to metallic state in heating
run, are nearly the same. In the cooling run the presence of time
dependence suggest that the switching happens somewhere between
$T_{MI}$ and $T^*$, stochastically. 
 Since we did not find any time dependence in heating run,
the switching temperature of an SR in a heating run is its local
$T_{MI}$. Thus the similarity in $\rho$ vs $1/T$ curves in heating
and cooling runs of the film suggest that $T^*$ and $T_{MI}$ are
very close for most of the SR's. The close value of $T^*$ and
$T_{MI}$ decreases the temperature range where an SR can exist in
its SC state in the cooling runs. This will result in a decrease
in the fraction of SR's present in their SC state compared to the
bulk. Thus we will have a smaller number of  SR's available to
switch from SC metallic state to insulating state with time and
hence we will get a weak  time dependence and a small hysteresis
in thin films which is in accordance with the observations. See
Figures \ref{fig:R vs T} and \ref{fig:time dep}.

The $T_{MI}$ and $T^*$ for most of the SR's in thin films are
nearly the same and thus the films have a small degree of
supercooling while  the bulk samples exhibit a high degree of
supercooling. This shows that energy of nucleation ($U$) for the
films is greatly reduced, which suggest that at the local level
the M-I transition in the films is occurring through heterogeneous
nucleation.
\subsection{Preferred sites for heterogeneous nucleation }
The preferential sites for nucleation are the free surface, the
thin film substrate interface, stacking faults, dislocation and
vacancies and in general it is found that the energy of nucleation
($U$) is minimum for free surface and is maximum for homogeneous
sites.\cite{Porter}. For the NdNiO$_3$ thin films on (001)
oriented NGO substrate, the probable places of nucleation are the
free surface of the film, the defects, and the film substrate
interface(because the high in-plane tensile strain of the
interface layers favor the existence of insulating state). The
formation of a critical nucleus of insulating phase at the defects
will have a smaller $\gamma_S$, the nucleation at the free surface
will have smaller $\gamma_S$ and $\Delta G_{Strain}$, and the
nucleation at interface will have a smaller $\Delta G_{Strain}$
compared to other regions of the films. Thus at a temperature $T$,
in the set of SR's that have the same M-I transition temperature,
those located at the free surface, the defects and the interface
will have a smaller energy of nucleation and this suggest that
these sites will act as preferred sites for heterogeneous
nucleation. Since our thin films have a high density of
heterogeneous nucleation sites (high concentration of defects,
large free surface area and favorable interface), the number of
SR's located at these sites will be high and these SR's will
transform from metallic to insulating state through  heterogeneous
nucleation. The transformation of an SR from the SC metastable
state to the insulating state occurs through a release of excess
energy.\cite{Porter} This excess thermal energy will be absorbed
by the nearby SC SR's and it may help them in forming the critical
nucleus of insulating phase. Further the surfaces of already
transformed SR's  have the same structure as that of critical
nucleus, their surfaces will partially reduce the surface energy
($\gamma_S$) required for formation of a critical nucleus in the
adjacent SC SR's. Thus the transformed SR's at the heterogeneous
nucleation sites will act as preferred sites for the  adjacent
SR's and so on.

  For Sm$_x$Nd$_{1-x}$NiO$_3$ thin films, an increase in the Sm
content increases the in-plane tensile strain and out-of-plane
compressed stress. The resistivity measurements on these samples
have shown that increase in Sm content broadens the M-I transition
and decreases the hysteresis.\cite{Ambrosinia} The reason for such
behavior was not clear and now it can be understood using the
concepts developed in this work. On increasing the Sm content, the
elastic strain energy of the film will increase and the film will
try to minimize it by increasing the number of defects. This
results in varying stress field and will cause a local variation
of $T_{MI}$ inside the films and thus the broadening of the M-I
transition. Further these enhanced defects sites will increase the
number of heterogeneous nucleation sites in the films which will
reduce the degree of supercooling of a fraction of SR's and it
will result in smaller hysteresis. Similar explanations can be
given for the case of Nd$_{0.7}$Eu$_{0.3}$NiO$_3$ thin films.

\section{Conclusion}
In conclusion our experimental results and discussion on NdNiO$_3$
thin films suggest that the physical properties of the films are
sensitive to deposition condition and film substrate lattice
mismatch strain. The elastic strain energy of the thin films is partially
relaxed by formation of defects which broadens the M-I transition of the films. %
The thin films and the bulk both exhibit non equilibrium features
in the transition regime which are attributed to the existence of
high temperature metallic phases in their supercooled metastable
state. The degree of supercooling in the thin films is found to be
much smaller than in the bulk samples whcih indicates that the M-I
transition in the bulk occurs through homogeneous nucleation while
the M-I transition in the films  occurs through heterogeneous
nucleation. Our analysis show that in comparison to the bulk
samples, the thin films have a higher density of heterogeneous
nucleation sites and hence the majority of the thin film
transforms from metallic to insulating state through heterogeneous
nucleation. Using these ideas we have also explained the higher
broadening and decrease in hysteresis in M-I transition of
Sm$_x$Nd$_{1-x}$NiO$_3$ and Nd$_{0.7}$Eu$_{0.3}$NiO$_3$ thin
films.

\section{Acknowledgements}
DK thanks the University Grants Commission of India for financial
support. RCB acknowledges the research grant from Indo-French
center for promotion of Advanced Research. 


\begin{thebibliography}{10}
\bibitem{endnote2} By the phrase 'complete metallic phase' we mean
a region which switches from metallic to insulating state as a
whole. It can be a crystallite or a portion of that (for more
detail see Ref.\ \onlinecite{Devendra}).
\bibitem{endnote} From FESEM images of the bulk sample and thin films, a rough estimate of the grain size
at the surface can be made. It is about 250~nm for bulk samples
and about 50~nm for thin films. This roughly agrees with the
values obtaine from X-ray peak width using Scherr formula.
\bibitem{Endnote} The details of measurement procedure is given in Ref.\
\onlinecite{Devendra}.
\bibitem{Endnote1} The magnitude of time dependence in resistivity $\rho(t,T)/\rho_0(0,T)$
and its variation with temperature depends on the volume fraction
$V$ of SC metallic SR's, their energy of nucleation $U$,  and the
temperature of percolation threshold. The resistivity of the
system is connected by the metallic volume fraction by McLachlan
GEM theory. We will get a higher timedependence in the volume when
the volume fraction of SC SR's having their $T^*$ within a few
Kelvin of the temperature of measurement ($T$) is high. The
detailed calculations for the bulk NdNiO$_3$ has been reported in
Ref.\ \onlinecite{Devendra}.
\end{thebibliography}
\end{document}